\documentclass[12pt, draftclsnofoot, journal, letterpaper, onecolumn]{IEEEtran}

\usepackage[dvips]{graphicx}
\usepackage{times}
\usepackage{cite}
\usepackage{amsmath}
\usepackage{array}
\usepackage{amssymb}
\usepackage{stfloats}
\usepackage{bm}
\usepackage{enumerate}
\usepackage{color}

\begin{document}

%
\title{Power and Subcarrier Allocation for Physical-Layer Security
in OFDMA-based Broadband Wireless Networks}

\author{Xiaowei Wang, Meixia Tao, Jianhua Mo and Youyun Xu
\thanks{The authors are with the Department of Electronic Engineering, Shanghai Jiao Tong University,
Shanghai, P.R. China 200240. Youyun Xu is also with the Institute of Communication Engineering,
PLA University of Science \& Technology, Nnjing, P.R. China 210007. Email: \{wangxiaowei, mxtao,
mjh\}@sjtu.edu.cn, yyxu@vip.sina.com.cn. }}

\maketitle
\begin{abstract}
Providing physical-layer security for mobile users in future broadband
wireless networks is of both theoretical and practical importance. In
this paper, we formulate an analytical framework for resource allocation
in a downlink OFDMA-based broadband network with coexistence of
secure users (SU) and normal users (NU). The SU's require secure data
transmission at the physical layer while the NU's are served
with conventional best-effort data traffic.
%
%
The problem is formulated as joint power and subcarrier allocation with
the objective of maximizing average aggregate information rate of all
NU's while maintaining an average secrecy rate for each individual
SU under a total transmit power constraint for the base station.
We solve this problem in an asymptotically optimal manner using dual
decomposition. Our analysis shows that an SU becomes a candidate
competing for a subcarrier only if its channel gain on this subcarrier
is the largest among all
and exceeds the second largest by a certain threshold.
Furthermore, while the power allocation for NU's follows the conventional
water-filling principle, the power allocation for SU's depends on both
its own channel gain and the largest channel gain among others.
We also develop a suboptimal algorithm to reduce the computational cost.
Numerical studies are conducted to evaluate the performance of the
proposed algorithms in terms of the achievable pair of information
rate for NU and secrecy rate for SU at different power consumptions.

\end{abstract}


\setlength\arraycolsep{2pt}

\section{Introduction}
Security is a crucial issue in wireless systems due to the
broadcasting nature of wireless radio waves. It also attracts
increasing attention because of the growing demand of private data
transmission such as online transaction and personal medical
information. Traditionally, cryptography undertakes most of security
work on upper layers, which is based on computational complexity. In
the standard five-layered protocol stack, security approaches are
designed on every layer except physical layer. Thus, establishing
physical-layer security is of both theoretical and practical
significance. In this study, we aim to provide physical-layer security
for mobile users in future broadband wireless networks and formulate
an analytical strategy for resource allocation to achieve this goal.

Information-theoretic security provides possibility of secure
transmission in the physical layer. By exploring secrecy capacity
and coding technique, messages can be sent without being decoded by
any eavesdropper. Information-theoretic security originates from
Shannon's notion of
perfect secrecy \cite{shannon}. He presented the general
mathematical structure and properties of secrecy systems. The
concept of information-theoretic security and wire-tap channel was
defined by Wyner \cite{wyner} and later by Csisz\'{a}r and
K\"{o}rner \cite{csiszar}, who proved the existence of channel
coding that makes the wiretapper to obtain no information about the
transmitted data. Then the study on information-theoretic security
is extended to various kinds of channels. Leung \emph{et al}
\cite{leung} focused on Gaussian wire-tap channel and showed that
secrecy capacity is the difference between the capacities of the
main and wire-tap channels. Barros \emph{et al} \cite{barros}
studied secrecy capacity in slow fading channels and introduced
outage into secrecy issues for the first time. In \cite{zangli},
Li \emph{et al} investigated independent parallel channels and
proved that the secrecy capacity of the system is the summation of
the secrecy rate achieved on each independent channel.
More recently, Zhu \emph{et al} \cite{zhu} studied the cooperative
power control for secret communications by using artificial noise
in symmetric Gaussian interference channel.

Orthogonal frequency division multiple access (OFDMA) has evolved
as a leading technology in future broadband wireless networks,
such as 3GPP Long-Term Evolution (LTE) and IEEE 802.16 WiMAX. It
enables efficient transmission of a wide variety of data traffic
by optimizing of power, subcarrier or bit allocation
among different users. In the past decade, tremendous research
results have been reported on the resource allocation of OFDMA
downlink networks, such as \cite{wong,Jang_JSAC03,Tao,
hui07,mokari,Song_TW05}, where the problem
formulation differs mostly in optimization objectives and
constraints. The work in \cite{wong} appeared as one of the earliest
results on margin adaptation for minimizing the total transmit power
with individual user rate requirements. Jang and Lee in \cite{Jang_JSAC03}
studied rate adaptation for system sum-rate maximization subject to
a total transmit power constraint. In \cite{Tao}, Tao \emph{et al} considered a
heterogenous network and studied the power and subcarrier allocation
problem for maximizing the sum-rate of non-delay-constrained users
while satisfying the basic rate requirement of each delay-constrained
user under a total transmit power constraint. Cross-layer optimizations
considering utility-function and traffic arrival distribution
for OFDMA networks were also studied in several works, e.g. \cite{hui07,mokari,Song_TW05}.
Nevertheless, none of these works take into account the security issue,
which attracts increasing attention recently in wireless networks
as aforementioned.

Secrecy or private message exchanges
between mobile users and base station (BS) are generally needed in present and
future wireless systems. Hence, it is essential to consider the
security demand when assigning radio resources to all users.
In \cite{jorswieck}, the authors made the initial attempt to find the
power and subcarrier allocation
in an OFDM-based broadcast channel with the objective
of maximizing sum secrecy rate. However, this work is confined
to two users only and does not consider the coexistence of
other types of users.

In this study,
we introduce two types of users according to
their secrecy demands in downlink OFDMA based broadband networks.
The first type of users have physical-layer security requirements and should be served
at a non-zero secrecy rate. These users are referred to
as \emph{secure users} (SU). The other type of users have no confidential
messages and do not care about security issues. Their traffic
is treated in a best-effort way. These users are regarded as
\emph{normal users} (NU).
{\color{blue} All SU's and NU's are legitimate users in the network and have their own data transmission with the BS. They are completely honest and always feedback the correct channel condition to the BS. Moreover, each of them is equipped with single antenna and only passively listens, rater than actively attacking using, for example, multiple antennas or colluding, when being a potential eavesdropper of the confidential messages for SU's.
The aim of this study is to investigate the power and subcarrier
allocation problem in such an OFDMA broadband network where the BS needs
to simultaneously serve multiple SU's and multiple NU's.}

The resource allocation problem in this paper possesses major differences compared
with those without secrecy constraint, owing to the coexistence of SU's and NU's.
Firstly, the legitimate subcarriers to be assigned
to an SU can only come from the subcarrier set on which this SU has the best
channel condition among all the users. This is because for each SU, any other
user in the same network is a potential eavesdropper. The authors in \cite{yugd}
showed that the secrecy capacity of a fading channel in the presence of multiple
eavesdroppers is the difference between the capacities of the main channel and the
eavesdropper channel with the largest channel gain among all eavesdroppers. As a
result, a non-zero secrecy rate on a subcarrier is possible to achieve only if the
channel gain of the SU on this subcarrier is the largest among all the users. Note that
this observation is very different from that in conventional OFDMA networks where a
user is still able to occupy some subcarriers and transmit signals over them even if
its channel gains on these subcarriers are not the largest. Secondly, even if an SU has
the best channel condition on a given subcarrier, assigning this subcarrier to the SU
may not be the optimal solution from the system perspective when the quality of service
of NU's is taken into account. This is due to the fact that the achievable secrecy rate
of channels with Gaussian noise is the subtraction of two logarithmic functions \cite{leung}, i.e.
\begin{eqnarray}
C_s = \left[\log(1+P\alpha_M) - \log(1+P\alpha_E)\right]^+
\end{eqnarray}
where  $P$ is the transmit power, $\alpha_M$ and $\alpha_E$ are the
channel-to-noise ratios (CNR) of the main channel and eavesdropper channel,
respectively, and $[x]^+=\max\{0,x\}$. In the case of large $P$, if
the gap between $\alpha_M$ and $\alpha_E$ is not large enough, the
achievable secrecy rate can be rather low. In this case it may be
more beneficial for the system to assign the subcarrier to an NU for
transmitting non-confidential data traffic.

Our goal is to find an optimal power and subcarrier allocation
policy to maximize the long-term aggregate information rate of all NU's while
maintaining a target average secrecy rate of each individual SU
under a total power constraint. Since a non-zero
instantaneous secrecy rate for each SU cannot be
guaranteed all the time due to channel fading, we assume that
the transmission of confidential
messages can wait until the channel condition of the SU becomes
favorable. As a result, the average secrecy rate
requirement instead of instantaneous secrecy rate requirement
is considered for each SU.
In addition, to make our problem more complete, both long-term average
and peak total transmit power constraints at the BS are considered.

Finding the optimal power and subcarrier allocation policy with
respect to channel conditions of all users in the considered system
is a functional optimization problem.
We solve this problem in dual domain using the decomposition method
in an asymptotically optimal manner. In particular, the joint
subcarrier assignment and power allocation at given dual variables can be
determined on a per-subcarrier basis. Our analysis shows that for
each subcarrier, if the user with the largest channel gain belongs to the category of NU's,
then this subcarrier will be assigned to this NU.
Otherwise if it is an SU, then this subcarrier will only
be assigned to this SU when its channel gain exceeds the
second-largest channel gain by a certain threshold. This analytical
finding agrees with the intuition mentioned above. It is also shown
that while the power allocation across the subcarriers assigned for NU's
follows the standard water-filling principle, the power allocation for SU's
depends on both the SU's channel gain and the second-largest channel gain.
Based on the insight derived from the
optimal policy, we further propose a low-complexity suboptimal
power and subcarrier allocation algorithm.
Firstly, we assign resources to only SU's as if all
the NU's were pure eavesdroppers without data traffic. Then,
the residual resources are allocated to NU's. The underlying mechanism of
this algorithm is to decouple the joint update of the Lagrangian
multipliers as required in the optimal algorithm so as to acquire a linear
complexity in the numbers of users and subcarriers.
%

The rest of the paper is organized as follows.  Section II describes
the system model and problem formulation. In Section III, we
introduce dual decomposition method to solve this problem under average
power constraint. The same problem under peak power constraint
is solved in Section IV. An efficient suboptimal resource allocation
algorithm is proposed in Section V. Section VI presents the performance
of the proposed scheme via simulation. Finally, we conclude this paper
in Section VII.

\section{System Model and Problem Formulation}

We consider the downlink of an OFDMA broadband network
with one BS and $K$ mobile users. The first $K_1$ users,
indexed as $k=1, \ldots, K_1$, are SU's. Each of them has
confidential messages to communicate with the BS and therefore
demands a secrecy rate no lower than a constant $C_k$, for $1\le k
\le K_1$. The other $K-K_1$ users, indexed as $k=K_1+1, \ldots, K$,
are NU's and demand service of best-effort traffic.
NU's and SU's are all assumed to eavesdrop the legitimate channel
non-cooperatively and each has only one antenna.
The communication link between the BS and each user is modeled as a
slowly time-varying frequency-selective fading channel. The channel
coefficients remain approximately unchanged during each time frame,
but vary from one frame to another in a random manner. The total
bandwidth is logically divided into $N$ orthogonal subcarriers by
using OFDMA with
each experiencing slow fading. As a central controller, the BS
knows the channel information of all users, finds the
allocation policy and then assigns power and subcarriers to mobile users
at each transmission frame according to the instantaneous
channel state information (CSI) of all users. The total transmit power
of the BS is subject to either a long-term average or peak constraint,
both denoted as power constraint $P$. {\color{blue}We assume full statistical
knowledge and instantaneous knowledge of CSI at the BS} and that
each subcarrier is occupied only by one user at each time frame to avoid
multi-user interference.

Let $\alpha_{k,n}$ denote the
CNR of user $k$ on subcarrier $n$ for all $k$ and $n$. The
system channel condition is denoted by the set
$\bm\alpha=\{\alpha_{k,n}\}$, which have a joint probability density
function of $f(\bm\alpha)$.
Let $\bm\Omega(\bm\alpha)=\{\Omega_1,...,\Omega_K\}$ denote the subcarrier
assignment policy with respect to the system channel condition
$\bm\alpha$, where $\Omega_k$ represents the set of subcarriers
assigned to user $k$. Furthermore, let
$\bm p(\bm\alpha)=\{p_{k,n},\forall k,\forall n\}$ denote the corresponding
power allocation policy,
where $p_{k,n}$ represents the transmit power allocated
to user $k$ on subcarrier $n$. We next present the achievable secrecy rate of SU
and the achievable information rate of NU separately, under a given resource
allocation policy $\{\bm p(\bm\alpha), \bm\Omega(\bm\alpha)\}$.

For SU $k, 1\leq k\leq K_1$, since the subcarriers are parallel to each other,
the achievable secrecy rate at a given channel realization is the summation
of those achieved on each subcarrier in the presence of $K-1$ potential
eavesdroppers, and can thus be expressed as
\cite{yugd}:
\begin{eqnarray}\label{eqn:cs}
r_k^s = \sum_{n \in \Omega_k} r_{k,n}^s
\end{eqnarray}
where
\begin{eqnarray}\label{eqn:csn}
r_{k,n}^s=\left[\log(1+p_{k,n}\alpha_{k,n})
-\log(1+p_{k,n}\beta_{k,n})\right]^+.
\end{eqnarray}
Here $\beta_{k,n} = \max\limits_{k', k' \ne k}
\alpha_{k',n}$ denotes the largest CNR among all the users except
user $k$ on subcarrier $n$. The expression (\ref{eqn:csn}) means that
non-zero instantaneous secrecy rate for SU $k$ on a subcarrier is
possible to achieve only if
its CNR on this subcarrier is the largest among all the $K$ users.
On the other hand, the achievable information rate of NU
$k$ for $K_1 < k \le K$ is given by:
\begin{eqnarray}\label{eqn:normalrate}
r_k=\sum_{n \in \Omega_k} r_{k,n},
\end{eqnarray}
where
\begin{eqnarray}\label{eqn:rn}
r_{k,n} =\log(1+p_{k,n}\alpha_{k,n}).
\end{eqnarray}
The problem is to find the optimal power and subcarrier
allocation policies $\{\bm{p}(\bm\alpha),(\bm\Omega(\bm\alpha)\}$ so
as to maximize the average aggregate information rate of the
$K-K_1$ NU's while satisfying the individual average secrecy
rate requirement for each of the $K_1$ SU's. We consider both
peak and average power constraints. This functional
optimization problem can be expressed as
\begin{eqnarray}\label{eqn:opt_p}
\max_{\{\bm\Omega({\boldsymbol\alpha}), {\bm p}({\boldsymbol\alpha})\}}
&& {\mathbb E}\left(\sum\limits_{k=K_1+1}^K \omega_k \sum_{n\in\Omega_k} r_{k,n}\right) \label{eqn:opt_obj} \\
{\rm{subject~to}}&&{\mathbb E}\left(\sum_{n\in\Omega_k} r_{k,n}^s\right)\geq C_k,\ 1\le k\le K_1 \label{eqn:opt_scon}\\
     &&\sum\limits_{k=1}^K \sum_{n\in\Omega_k} p_{k,n} \leq P \label{eqn:opt_pcon_in} \\
  or &&{\mathbb E}\left(\sum\limits_{k=1}^K \sum_{n\in\Omega_k} p_{k,n}\right)\leq P \label{eqn:opt_pcon_av} \\
     &&p_{k,n}\geq 0, \forall k,n \label{eqn:opt_pge0}\\
     &&\Omega_1\cup...\cup\Omega_K\subseteq \{1,2,...,N\} \nonumber\\
     &&\Omega_1,...,\Omega_K \;{\rm are~ disjoint}
     \label{eqn:opt_subc}
\end{eqnarray}
where notation ${\mathbb E}$ represents statistical
average over the joint distribution of channel conditions, i.e.
${\mathbb E}[\cdot] = \int{(\cdot)f(\bm\alpha)d\bm\alpha}$,
and $\omega_k$ is a weighting parameter of NU $k$, representing its quality-of-service demand.
Constraint (\ref{eqn:opt_pcon_in}) is the peak total power constraint while
(\ref{eqn:opt_pcon_av}) is the average total power constraint.

In the above formulation, we impose the long-term average secrecy rate constraints
because instantaneous non-zero secrecy rate cannot be guaranteed at every frame.
It is possible due to channel fading that in a certain frame,
an SU is the best user on none of the $N$ subcarriers and thus obtains a zero secrecy
rate. Therefore, power and subcarrier allocation should be adapted every frame
to meet a long-term secrecy rate requirement.

\section{Optimal Resource Allocation Under Average Power Constraint}
In this section, we solve the problem with the average power constraint
formulated above. It is not difficult to observe that this
problem satisfies the time-sharing condition introduced in \cite{weiyu}.
That is, the objective function is concave and constraint (\ref{eqn:opt_scon}) is
convex given that $r_{k,n}^s$ is concave in $p_{k,n}$ and that the
integral preserves concavity. Therefore, similar to the OFDMA networks
without secrecy constraint, we can use dual approach for resource
allocation and the solution is asymptotically optimal for large enough
number of subcarriers.

Define $\mathcal{P}({\boldsymbol\alpha})$ as a set of all possible
non-negative power parameters $\{p_{k,n}\}$ at any given system
channel condition $\boldsymbol\alpha$ satisfying that for each
subcarrier $n$ only one $p_{k,n}$ is positive.
This definition takes into account both the power constraint
(\ref{eqn:opt_pge0}) and the exclusive subcarrier allocation
constraint (\ref{eqn:opt_subc}).
The Lagrange dual function is thus given by
\begin{eqnarray} \label{eqn:dual}
&&g(\bm\mu, \lambda) \nonumber\\
&&= \max\limits_{\{p_{k,n}\}\in\mathcal{P}(\bm\alpha)}
\Bigg\{ \mathbb{E} \left[ \sum_{k = K_1 +1}^K \omega_k
\sum_{n = 1}^N r_{k,n}\big(p_{k,n}(\bm\alpha),\bm\alpha\big)\right]\nonumber\\
&&~~+\sum\limits_{k = 1}^{K_1}{\mu _k  \left( \mathbb{E}\left[\sum_{n = 1}^N r_{k,n}^s
\big(p_{k,n}(\bm\alpha),\bm\alpha\big)\right] - C_k \right)}
+\lambda \left(P-\mathbb{E}\left[\sum_{k = 1}^K \sum\limits_{n= 1}^N {p_{k,n}(\bm\alpha)}\right] \right) \Bigg\},
\end{eqnarray}
where
$\bm\mu=(\mu_1, \ldots, \mu_{K_1})\succeq 0$ and $\lambda \geq 0$ are the
Lagrange multipliers for the constraints (\ref{eqn:opt_scon}) and
(\ref{eqn:opt_pcon_av}) respectively, and notation $\mathbb{E}$ stands
for the statistical average over all channel conditions $\bm\alpha$.
Then the dual problem  of the original problem (\ref{eqn:opt_p}) is
given by
\begin{eqnarray}\label{eqn:ming}
&&\min \ g(\bm{\mu}, \lambda)\nonumber\\
&&s.t. \ \bm{\mu}\succeq 0, \lambda \geq 0.
\end{eqnarray}

Observing (\ref{eqn:dual}), we find that the maximization in the
Lagrange dual function can be decomposed into $N$ independent sub-functions as:
\begin{eqnarray}\label{eqn:g}
g(\bm{\mu}, \lambda)= \sum\limits_{n = 1}^N g_n(\bm{\mu}, \lambda)
-\sum\limits_{k = 1}^{K_1}{\mu_k C_k}+\lambda P,
\end{eqnarray}
where
\begin{eqnarray}\label{eqn:gnmax}
g_n(\bm{\mu}, \lambda)&=&\max_{\{p_{k,n}\}\in
\mathcal{P}(\bm\alpha)}\mathbb{E}\left[ J_n\left(\bm\mu, \lambda,
\bm\alpha, \{p_{k,n}\}_{k}\right)\right],
\end{eqnarray}

with
\begin{eqnarray} \label{eqn:Jna}
  J_n(\bm\mu, \lambda, \bm\alpha, \{p_{k,n}\}_{k}) =
  \sum_{k=K_1+1}^K \omega_k r_{k,n}  + \sum_{k=1}^{K_1} \mu_k r_{k,n}^s -
     \lambda\sum_{k=1}^K p_{k,n}.
\end{eqnarray}

For fixed $\bm{\mu}$ and $\lambda$, the maximization problem in
(\ref{eqn:gnmax}) is a single-carrier multiple-user power allocation
problem.
Given that the expectation $\mathbb{E}$ is over the channel
condition $\bm\alpha$, we can obtain the maximum by directly
maximizing the function (\ref{eqn:Jna}). Based on the above
sub-problems, we now discuss the optimality conditions of
power allocation and subcarrier assignment, respectively in
the following two subsections.

\subsection{Optimality Condition of Power Allocation}
The function in (\ref{eqn:Jna}) is concave in $p_{k,n}$ and hence
its maximum value can be found by using Karush-Kuhn-Tucker (KKT) conditions.
Specifically, suppose that subcarrier $n$ is assigned to user $k$,
then taking the partial derivation of $J_n(\bm\mu, \lambda,
\bm\alpha, \{p_{k,n}\}_{k})$ with respect to $p_{k,n}$ and equating it
to zero, we obtain the following optimality condition of power allocation:
\begin{eqnarray}\label{psu}
p_{k,n}^*=\frac{1}{2} \Bigg[ \sqrt{\left(\frac{1}{\alpha_{k,n}}-\frac{1}{\beta_{k,n}}\right)^2
+\frac{4\mu_k}{\lambda}\left(\frac{1}{\beta_{k,n}}-\frac{1}{\alpha_{k,n}}\right)}
-\left(\frac{1}{\alpha_{k,n}}+\frac{1}{\beta_{k,n}}\right)
\Bigg]^+
\end{eqnarray}
for $k=1,\ldots, K_1$, and
\begin{eqnarray}\label{pNU}
p_{k,n}^*&=&\left[\frac{\omega_k}{\lambda}-\frac{1}{\alpha_{k,n}} \right]^+
\end{eqnarray}
for $k=K_1+1,\ldots,K$.

We can conclude from (\ref{pNU}) that the optimal power allocation
for NU's follows the conventional water-filling principle and the water level
is determined by both the weight of the NU and the average power
constraint. On the other hand, it is seen from (\ref{psu}) that the
optimal power allocation for SU's has the same form as the
result obtained in \cite{Gopala} for conventional fading wire-tap
channels, as expected. By observing (\ref{psu}) closely it is also seen
that the SU must satisfy $\alpha_{k,n}-\beta_{k,n} \geq
\frac{\lambda}{\mu_k}$ in order to be allocated non-zero power.
This means that the power allocation for SU depends on both the channel
gain of the SU and the largest channel gain among all the other users.
Moreover, it is non-zero only if the former exceeds the latter by the
threshold $\frac{\lambda}{\mu_k}$.

\subsection{Optimality Condition of Subcarrier Assignment}
Next, substituting (\ref{psu}) and (\ref{pNU}) into
(\ref{eqn:gnmax}) and comparing all the $K$ possible user
assignments for each subcarrier $n$,  we obtain
\begin{equation}\label{eqn:gn_max}
   g_n(\bm\mu ,\lambda) = \mathbb{E} \left[ \max\limits_{1\leq k \leq K} H_{k,n}
   \left( \bm\mu,\lambda, \bm\alpha \right) \right],
\end{equation}
where the function $H_{k,n}(\cdot)$ is defined as
\begin{eqnarray}\label{eqn:H_su}
H_{k,n}\left( \bm\mu,\lambda, \bm\alpha \right)
&=& \mu_k\log\left(\frac{1+p_{k,n}^*\alpha_{k,n}}
{1+p_{k,n}^*\beta_{k,n}}\right)-\lambda p_{k,n}^*
\end{eqnarray}
for $1\le k \le K_1$ and $p_{k,n}^*$ defined in (\ref{psu}),
and
\begin{eqnarray}\label{eqn:H_nu}
H_{k,n}\left( \bm\mu,\lambda, \bm\alpha \right)
= \omega_k \left[\log\frac{\omega_k \alpha_{k,n}}{\lambda}\right]^+
-\left[ \omega_k -\frac{\lambda}{\alpha_{k,n}}\right]^+
\end{eqnarray}
for $K_1 < k \le K$.

From (\ref{eqn:gn_max}) it is observed that the function $H_{k,n}$
defined in (\ref{eqn:H_su}) and (\ref{eqn:H_nu}) plays an important
role in determining the optimal subcarrier assignment. In specific,
for any given dual variables $\bm\mu$ and $\lambda$, the subcarrier
$n$ will be assigned to the user with the maximum value of
$H_{k,n}$. That is, the optimality condition for subcarrier assignment
is given by
\begin{eqnarray}\label{opt_k}
k_n^*=\arg \max\limits_k H_{k,n},~~~{\rm for}~n=1,...,N.
\end{eqnarray}

Note that for $k=K_1+1,...,K$, $H_{k,n}$ is monotonically
increasing in $\alpha_{k,n}$. Therefore, the NU with larger $\alpha_{k,n}$
is more likely to be assigned subcarrier $n$.
We also notice that for $k=1,...,K_1$, $H_{k,n}>0$ only when SU $k$
has the largest $\alpha_{k,n}$ among all the $K$ users and satisfies
$\alpha_{k,n}>\beta_{k,n}+\lambda/\mu_k$. Otherwise, $H_{k,n}=0$.
In other words, an SU becomes a candidate
for subcarrier $n$ only if its CNR is the largest and is
$\lambda/\mu_k$ larger than the second largest.

\subsection{Dual Update}

Substituting $g_n(\bm\mu,\lambda)$ for $n=1,\ldots, N$ into
(\ref{eqn:g}), we obtain $g(\bm\mu, \lambda)$. As studied in
\cite{boyd}, the dual problem (\ref{eqn:ming}) is always convex and
can be minimized by simultaneously updating $(\bm\mu, \lambda)$
using gradient descent algorithms. However, note that $g(\bm\mu,
\lambda)$ is not differentiable due to the discontinuity of
subcarrier assignment and hence its gradient does not exist.
Nevertheless, we present a subgradient of $g(\bm\mu, \lambda)$ as
follows:
\begin{eqnarray}
\Delta\mu_k &=& \mathbb{E}\left[\sum\limits_{n = 1}^N r_{k,n}^{s*}\right] -C_k, \label{eqn:deltamu} \\
\Delta\lambda &=& P- \mathbb{E}\left[\sum_{k=1}^K \sum\limits_{n =
1}^N p_{k,n}^* \right], \label{eqn:deltalam}
\end{eqnarray}
where $r_{k,n}^{s*}$ is obtained by substituting the optimal
$p_{k,n}^*$ into (\ref{eqn:csn}).
A brief proof is as follows.

From the expression of $g(\bm\mu,\lambda)$ in (\ref{eqn:dual}), we get
\begin{eqnarray} \label{eqn:subg}
&&g(\bm\mu', \lambda') \nonumber\\
&&\geq  \mathbb{E}
\left[ \sum_{k = K_1 +1}^K \omega_k
\sum_{n = 1}^N r_{k,n}^*\right]
+\sum\limits_{k = 1}^{K_1}{\mu _k'
\left( \mathbb{E}\left[\sum_{n = 1}^N r_{k,n}^{s*}\right]
- C_k \right)}
+\lambda' \left(P-
\mathbb{E}\left[\sum\limits_{k = 1}^K \sum_{n= 1}^N
{p_{k,n}^*} \right]  \right) \nonumber\\
&&=g(\bm\mu, \lambda)
+\sum\limits_{k = 1}^{K_1}{(\mu _k'-\mu_k)
\left( \mathbb{E}\left[\sum_{n = 1}^N r_{k,n}^{s*}\right]
- C_k \right)}
+\left(\lambda'-\lambda\right)
\left(P-\mathbb{E}\left[\sum\limits_{k = 1}^K \sum_{n= 1}^N
{p_{k,n}^*} \right]  \right) .
\end{eqnarray}
The results in (\ref{eqn:deltamu}) and (\ref{eqn:deltalam})
are thus proved by the definition of subgradient.

In the case when the numerical
computation of statistical average is too complex, we can
change to time average when the channel fading process is
ergodic. Specifically,
\begin{eqnarray}
 \mathbb{E}\left[\sum\limits_{n = 1}^N r_{k,n}^{s*}\right] &=
 & \lim_{t\to\infty} \frac{1}{t} \sum_{t'=1}^t\sum\limits_{n = 1}^N r_{k,n}^{s*}(t'),\\
 \mathbb{E}\left[\sum_{k=1}^K \sum\limits_{n =
1}^N p_{k,n}^*\right]  &=& \lim_{t\to\infty}
\frac{1}{t}\sum_{t'=1}^t\sum_{k=1}^K \sum\limits_{n = 1}^N
p_{k,n}^*(t').
\end{eqnarray}

After finding the optimal dual variables $\{\mu_k^*\}$ and
$\lambda^*$, the optimal power and subcarrier allocation policy
is then obtained by substituting them into the optimality
conditions (\ref{psu}), (\ref{pNU}) and (\ref{opt_k}).

\subsection{Discussion of Feasibility}
To make the above optimization problem feasible, the secrecy
requirement $C_k$ for each SU must be chosen properly according to
the total power constraint $P$. In this subsection, we will
derive an upper bound of average secrecy rate each SU can obtain.
If $C_k$ is set equal or greater than this upper bound, the secrecy
requirement cannot be satisfied and thus this optimization
problem is not feasible.

For simplicity, we assume that the channel conditions of all the
$K$ users on each subcarrier are independently and identically distributed
and follow Rayleigh distribution. Then, each SU has a probability of
$1/K$ to be the best user on each subcarrier. As a result, we can write
the average achievable secrecy rate of each SU $k$ as
\begin{eqnarray}
\bar R_k^s=\frac{N}{K}\mathbb{E}[r_{k,n}^s].
\end{eqnarray}
Assuming that the transmit power goes to infinity, the maximum
per-subcarrier secrecy rate $r_{k,n}^s$ can be obtained as
\begin{align}\label{limr}
\lim_{p_{k,n} \to \infty }r_{k,n}^s
=\left[\log\frac{\alpha_{k,n}}{\beta_{k,n}}\right]^+ .
\end{align}
Therefore, an upper bound on $\bar R_{k}^s$ can be theoretically
derived using order statistics as
\begin{eqnarray}\label{bound_ck}
\bar R_k^s && \leq  \frac{N}{K} \int_0^\infty \int_{\nu_2}^\infty log\frac{\nu_1}{\nu_2}
f(\nu_1,\nu_2) d\nu_1 d\nu_2 \nonumber\\
&&=\frac{N}{K} \int_0^\infty  \int_{\nu_2}^\infty   N(N - 1){{(1 - {e^{ - \frac{\nu_2}{\rho }}})}^{N - 2}}
\frac{1}{\rho }e^{ - \frac{\nu_2}{\rho }}\frac{1}{\rho }e^{ - \frac{\nu_1}{\rho }}
\log\frac{\nu_1}{\nu_2}d\nu_1 d\nu_2  \nonumber \\
&&=\frac{N}{K} \int_0^\infty \int_{\nu_2}^\infty  N(N - 1){{(1 - {e^{ - \frac{\nu_2}{\rho }}})}^{N - 2}}
\frac{1}{\rho }e^{ - \frac{\nu_2}{\rho }}
\frac{1}{\rho }e^{ - \frac{\nu_1}{\rho }}\log \nu_1 d\nu_1 d\nu_2  \nonumber \\
&&~~ -\frac{N}{K}\int_0^\infty  \int_{\nu_2}^\infty  N(N - 1){(1 - e^{ - \frac{\nu_2}{\rho }})}^{N - 2}
\frac{1}{\rho }{e^{ - \frac{\nu_2}{\rho }}}
\frac{1}{\rho }e^{ - \frac{\nu_1}{\rho }}\log \nu_2 d\nu_1 d\nu_2.
\end{eqnarray}
In the above derivation, $\nu_1$ and $\nu_2$ denote the largest and second
largest CNR's on a given subcarrier, $f(\nu_1,\nu_2)$ is
the joint probability density function of $\nu_1$ and
$\nu_2$, which can be obtained through order statistics.
We assume Rayleigh fading and the probability distribution
function of CNR is $f(x)=\frac{1}{\rho}e^{-\frac{x}{\rho}}$,
where $\rho$ is the mean value.
Since it is difficult to further express the integrals in (\ref{bound_ck}) in
a closed form, we use numerical integration to get
the upper limit value of secrecy rate. As an example, when
the parameters are set to be $N=64$, $K=8$, $K_1=4$ and $\rho=1$,
we obtain $\bar R_k^s \leq$ 3.5nat/OFDMA symbol.
In this case, when the secrecy rate constraint $C_k > 3.5$,
the problem becomes infeasible.

\section{Optimal Resource Allocation under Peak Power Constraint}
In Section III, we solved the optimization problem under
long-term average power constraint. In this section,
we solve the same problem (\ref{eqn:opt_p}) but under peak
power constraint. Note that the peak power constraint is
more suitable for practical systems as the BS usually has a
maximum radiation power. By using the similar definition of
power parameter set $\mathcal{P}(\bm\alpha)$ as in the previous section,
the associated Lagrange dual function is given by
\begin{eqnarray} \label{eqn:dual2}
&& g(\bm\mu, \lambda(\bm\alpha))\nonumber \\
&&= \max\limits_{\{p_{k,n}\}\in
{\mathcal{P}}(\bm\alpha)} \Bigg\{ \mathbb{E} \left[ \sum_{k = K_1 +
1}^K \omega_k
\sum_{n = 1}^N r_{k,n}\big(p_{k,n}(\bm\alpha),\bm\alpha\big)\right]
+\sum\limits_{k = 1}^{K_1}\mu _k \left( \mathbb{E}\left[\sum_{n = 1}^N r_{k,n}^s
\big(p_{k,n}(\bm\alpha),\bm\alpha \big)\right] - C_k \right)\nonumber \\
&&~~~+\int_{\bm\alpha} \lambda(\bm\alpha) \left(P-
\sum_{k = 1}^K \sum\limits_{n= 1}^N {p_{k,n}(\bm\alpha)} \right)f(\bm\alpha) d\bm\alpha \Bigg\}.
\end{eqnarray}
The difference from the dual function with average power constraint
in (\ref{eqn:dual}) is that the dual variable $\lambda(\bm\alpha)$
associated with the power constraint (\ref{eqn:opt_pcon_av}) is a
function of the system channel condition $\bm\alpha$.

This dual function can be similarly decomposed into $N$
independent sub-functions with each given by
\begin{eqnarray}\label{eqn:gnmax2}
g_n(\bm\mu, \lambda(\bm\alpha))
=\max_{\{p_{k,n}\}\in
\mathcal{P}(\bm\alpha)} \mathbb{E} \left[ J_n\left(\bm\mu, \lambda(\bm\alpha),
\bm\alpha, \{p_{k,n}\}_{k}\right) \right],
\end{eqnarray}

with
\begin{eqnarray}\label{eqn:J2}
J_n\left(\bm\mu, \lambda(\bm\alpha),
\bm\alpha, \{p_{k,n}\}_{k}\right)
=\sum_{k=K_1+1}^K \omega_k r_{k,n}
+ \sum_{k=1}^{K_1} \mu_k r_{k,n}^s
-\lambda(\bm\alpha)\sum_{k=1}^K p_{k,n}.
\end{eqnarray}
Given that the order of expectation operation and the max operation
can be reversed, we can obtain $g_n(\bm\mu, \lambda(\bm\alpha))$
by maximizing the function (\ref{eqn:J2}) for every $\alpha_{k,n}$.

Suppose that subcarrier $n$ is assigned to user $k$.
Taking the partial derivative of $J_n(\bm\mu, \lambda(\bm\alpha),
\bm\alpha, \{p_{k,n}\}_{k})$ with respect to $p_{k,n}$ and equating it
to zero,  we obtain the optimality conditions of power allocation

\begin{eqnarray}\label{psu2}
p_{k,n}^*=\frac{1}{2} \Bigg[ \sqrt{\left(\frac{1}{\alpha_{k,n}}-\frac{1}{\beta_{k,n}}\right)^2+\frac{4\mu_k}{\lambda(\bm\alpha)}\left(\frac{1}{\beta_{k,n}}-\frac{1}{\alpha_{k,n}}\right)}
-\left(\frac{1}{\alpha_{k,n}}+\frac{1}{\beta_{k,n}}\right)
\Bigg]^+
\end{eqnarray}
for $k=1,\ldots, K_1$, and
\begin{eqnarray}\label{pNU2}
p_{k,n}^*=\left[\frac{\omega_k}{\lambda(\bm\alpha)}-\frac{1}{\alpha_{k,n}}\right]^+
\end{eqnarray}
for $k=K_1+1,\ldots,K$.

Comparing (\ref{psu2}) and (\ref{pNU2}) with (\ref{psu}) and (\ref{pNU}),
we observe that the optimal power allocations under the average power
constraint and the peak power constraint have similar structure.
The difference lies
in the dual variable that controls total power. For average
power constraint, $\lambda$ is a constant for all $\alpha_{k,n}$.
For peak power constraint, this variable
changes with $\{\alpha_{k,n}\}$.

The rest of the algorithm is alike to the problem in the previous
section. The difference is that the subgradient to update
$\lambda(\bm\alpha)$ is changed to
\begin{eqnarray}
\Delta\lambda(\bm\alpha) &=& P- \sum_{k=1}^K \sum\limits_{n =
1}^N p_{k,n}^*(\bm\alpha) .
\end{eqnarray}

\section{Suboptimal Power and Subcarrier Allocation Algorithm}
The complexity of the optimal power and subcarrier allocation
policy presented in the previous two sections mainly lies in
the joint optimization of Lagrange multipliers
$\bm\mu=(\mu_1, \ldots, \mu_{K_1})$ and $\lambda$. If we choose
the ellipsoid method, it converges in $\mathcal{O}((K_1+1)^2 \log\frac{1}{\epsilon})$
iterations where $\epsilon$ is the accuracy \cite{boyd}. Thus, if the number of SU
is large, the computational complexity may not be favorable for
practical implementation. Based on the insight derived from the
optimal power and subcarrier allocation policy in the previous
sections, we present in this section a low-complexity and
efficient suboptimal power and subcarrier allocation algorithm.
For simplicity, only the average power constraint is considered.
The extension to the peak power constraint is simple. The idea
of this scheme is to first assign the resources to only SU's as
if all the NU's were pure eavesdroppers without data transmission.
After that, the residual subcarriers and power, if any,
are distributed among NU's. By doing this, the joint update of
the Lagrange multipliers will be decoupled as detailed below.

In this scheme, the power allocation adopts the expressions in
(\ref{psu}) and (\ref{pNU}) except that the parameter
$\lambda/\mu_k,k=1,\ldots,K_1$ in (\ref{psu}) is replaced by a
new variable $\nu_k$.
Also, in (\ref{pNU}) we define $L_k={\omega_k}L_0$, for $k=K_1+1,\ldots,K)$
where $L_0=\frac{1}{\lambda}$. The power allocation scheme among
NU's is water-filling with different water levels which are
proportional to NU's weights. Through simple observation, for SU
$p_{k,n}$ is monotonically decreasing in $\nu_k$ and thus $r_{k,n}^{s}$
is monotonically decreasing in $\nu_{k}$. Intuitively, for NU $p_{k,n}$
is monotonically increasing in water level $L_0$. Therefore,
$\nu_{k}$ and $L_0$ can be found through two separate binary searches.
We use a set of training channel realizations to compute the statistical
average numerically. The set is sufficiently large to assure that
their distribution converges to the statistical distribution.

The outline of this suboptimal algorithm is presented below.

\vspace{0.1cm}\hrule\vspace{0.3cm}
\textbf{Suboptimal Algorithm}
\vspace{0.1cm} \hrule \vspace{0.3cm}
~~\textbf{Find the optimal $\nu_k$ to achieve the secrecy rate requirement $C_k$ for $k=1,\ldots,K_1$.}
\begin{enumerate}
\item
Set $\nu_k^{UB}$ sufficiently large, $\nu_k^{LB}=0$ and $\nu_k=\frac{1}{2}(\nu_k^{LB}+\nu_k^{UB})$.
\item
For every training channel realization $\bm\alpha$

~~~Find $\Omega_k=\{n:\alpha_{k,n}>\max_{k'\neq k}(\alpha_{k',n})+\nu_k\}$;

~~~Compute $p_{k}(\bm\alpha)=\sum\limits_{n\in\Omega_k}
p_{k,n}(\bm\alpha)$ and
$r_{k}^s(\bm\alpha)=\sum\limits_{n\in\Omega_k} r_{k}^s(\bm\alpha)$,
where $p_{k,n}(\bm\alpha)$ and $r_{k,n}^{s}(\bm\alpha)$ are computed
according to (\ref{psu}) and (\ref{eqn:csn}), respectively with
$\lambda/\mu_k$ replaced by $\nu_k$;

Compute $\overline{r}_k^s=\mathbb{E}\left[r_{k}^{s}(\bm\alpha)\right]$
and $\overline{p}_{k}=\mathbb{E}\left[p_{k}(\bm\alpha)\right]$.
\item
If $\overline{r}_k^s>C_k$, $\nu_k^{LB}$ $=\nu_k$,
else $\nu_k^{UB}$ $=\nu_k$. Set $\nu_k=\frac{1}{2}(\nu_k^{LB}+\nu_k^{UB})$.
\item
Repeat Steps 2)-3) until $|\overline{r}_k^s-C_k|\leq\epsilon C_k$, for each $k\in[1,K_1]$ .
\item
Compute the power consumed by SU's, $\overline{P}_{SU}=\sum\limits_{k=1}^{K_1}\overline{p}_k$.
\vspace{0.1cm} \hrule \vspace{0.3cm}
\textbf{Find the optimal water level $L_0$ to meet the power constraint.}
\item
Set $L_0^{UB}$ sufficiently large, $L_0^{LB}=0$ and $L_0=\frac{1}{2}(L_0^{UB}+L_0^{LB})$.
\item
For every training channel realization $\bm\alpha$

~~~For every residual subcarrier $n \notin \bigcup\limits_{k=1}^{K_1} \Omega_k$, i.e. not occupied by SU's

~~~~~~Find $k=\arg\max\limits_{k\in(K_1,K]}{H_{k,n}}$ according to (\ref{eqn:H_nu}) with $1/\lambda$ replaced by $L_0$;

~~~~~~For the found $k$, compute $p_{k,n}(\bm\alpha)$ and $r_{k,n}(\bm\alpha)$ according to (\ref{pNU}) and (\ref{eqn:rn})
       with $1/\lambda$ replaced by $L_0$;


Compute $\overline{r}_k=\mathbb{E}\left[\sum\limits_{n = 1}^N r_{k,n}(\bm\alpha)\right]$ and
$\overline{P}_{NU}=\mathbb{E}\left[\sum\limits_{k=K_1+1}^{K} \sum\limits_{n=1}^N p_{k,n}(\bm\alpha)\right]$.
\item
If $\overline{P}_{NU}>P-\overline{P}_{SU}$, $L_0^{UB}=L_0$, else
$L_0^{LB}=L_0$. Set $L_0=\frac{1}{2}(L_0^{UB}+L_0^{LB})$.
\item
Repeat Steps 7)-8) until $|\overline{P}_{SU}+\overline{P}_{NU}-P|<\epsilon P$.
\end{enumerate}

\vspace{0.1cm} \hrule \vspace{0.3cm}

In the algorithm, we first find the the subcarrier set
$\Omega_k(k=1,\ldots,K_1)$ of SU's. The criterion is whether SU's
CNR is $\nu_k$ larger than the second largest. As $\Omega_k$'s
are disjoint, the optimal $\nu_k$ satisfying $C_k$ can be
obtained through $K_1$ independent binary searches. After
getting $\{\nu_k\}$, the power allocation for the $K_1$
SU's is determined and the total power left for $K-K_1$ NU's is also known.
The water levels of power allocation (\ref{pNU}) for NU's
can be searched until the rest of the total power for NU's is used up.

Since $\nu_k(k=1,\ldots,K_1)$ and $L_0$ are obtained individually by
binary search, this suboptimal algorithm converges in
$\mathcal{O}(K_1 \log\frac{1}{\epsilon}+\log\frac{1}{\epsilon})=\mathcal{O}((K_1+1)\log\frac{1}{\epsilon})$
iterations. Note that the optimal algorithm converges in
$\mathcal{O}((K_1+1)^2 \log\frac{1}{\epsilon})$ iterations as mentioned in
the beginning of this section.
In addition, the computational loads in each iteration of both the
optimal and suboptimal schemes are linear in $KN|\bm\alpha|$ where $|\bm\alpha|$
is the number of the training channel realizations. So the suboptimal
scheme reduces the complexity by about $\frac{1}{K_1+1}$.

\section{Numerical Results}
In this section, we provide some numerical results to evaluate the performance of
proposed optimal and suboptimal resource allocation algorithms
under both average and peak power constraints. In the simulation setup, we consider
an OFDMA network with $N=64$ subcarriers and $K=8$ mobile users, among
which $K_1=4$ are SU's and $K-K_1=4$ are NU's. For simplicity, all the
weighting parameters ${\omega_k}$'s for NU's are set to 1 and the secrecy rate requirements
$C_k$'s for SU's are set to be identical, denoted as $C_k=R_{SU}$.
Let $R_{NU}$ denote the average total information rate of the NU's.
Here we use nat instead of bit as the measurement unit of data  for
computational convenience. {\color{blue}The channel on each subcarrier
for each user is assumed to be independent and identically distributed
Rayleigh fading with unit mean-square value for illustration purpose only.}
Note that more sophisticated multi-path broadband channel models, such as HiperLan/2 channel model A, can be easily applied since our analytical framework is general and applicable to arbitrary channel distributions. The system total transmit SNR defined in the simulation is equivalent to the average power budget $P$ (or peak power budget if the peak power constraint is concerned) on all the $N$ subcarriers at the base station, assuming unit noise power.

To evaluate the optimal and suboptimal power and subcarrier adaptation schemes,
we introduce two non-adaptive
schemes in this simulation as benchmarks. In these two
non-adaptive schemes, subcarrier assignment is fixed beforehand while power
allocated to the predetermined subcarrier sets conforms to (\ref{psu}) and
(\ref{pNU}). In the first fixed subcarrier assignment scheme, denoted as
FSA-1, the 64 subcarriers are equally assigned to the 8 users regardless
of user type and thus each user obtains 8 subcarriers. In the second
scheme, denoted as FSA-2, the SU's are given higher priority and each
is assigned 12 subcarriers, whereas each NU is assigned 4 subcarriers.

\begin{figure}[!hbt]
\begin{centering}
\includegraphics[scale=0.8]{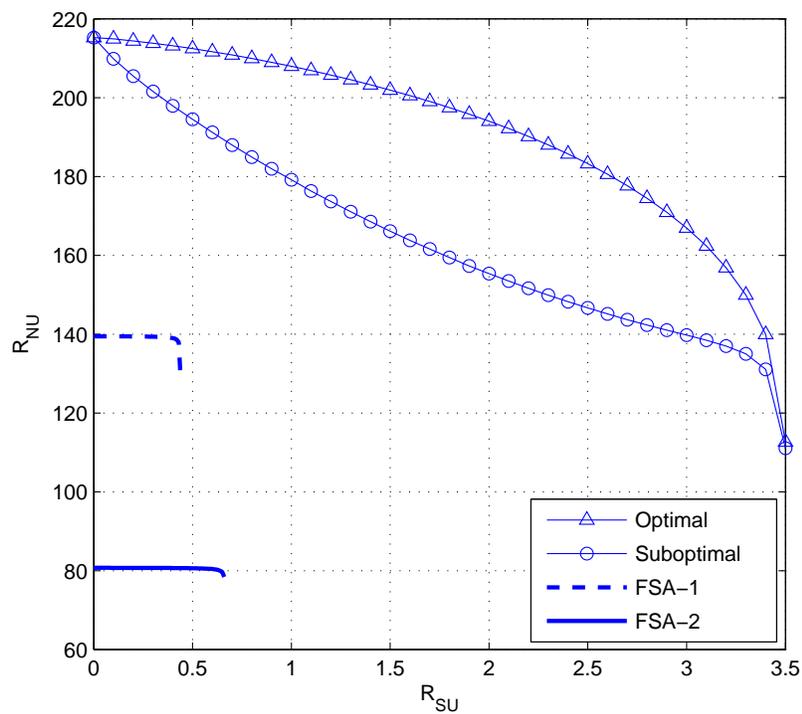}
\vspace{-1em}
 \caption{Achievable $(R_{SU},R_{NU})$ pair at total transmit SNR of 30dB.} \label{fig:1}
\end{centering}
\vspace{-0.1em}
\end{figure}

We first demonstrate the pair of achievable average total
information rate of NU's and feasible average secrecy rate
requirement of each individual SU $(R_{NU},R_{SU})$ at a given
average total power constraint. Fig.~\ref{fig:1} shows the rate pair
at fixed total transmit SNR=30dB. First it is observed that using
both optimal and suboptimal algorithms, $R_{NU}$ decreases with the
increase of secrecy requirement $R_{SU}$ and falls sharply to zero
at around $R_{SU}=3.5$ nat/OFDM symbol. Recall the discussion in
Section III-D where an upper bound on the secrecy rate requirement
to make the problem feasible was obtained as $3.5$ nat/OFDM symbol
in the case of $N=64$ subcarriers and $K=8$ users. This explains the
drastic fall of $R_{NU}$.
It is then observed from Fig.~\ref{fig:1} that the suboptimal algorithm
only incurs less than $20\%$ loss in $R_{NU}$ when achieving the same $R_{SU}$
compared with the optimal algorithm.
Now comparing the optimal and suboptimal schemes with the
non-adaptive ones, both of them earn great advantage over the two
benchmarks FSA-1 and FSA-2. In particular, the maximum feasible points
of the two fixed subcarrier assignment methods appear at around
$R_{SU}=0.44$ and $0.66$ nat/OFDM symbol, respectively, which are
much lower than those in the optimal and suboptimal algorithms.

\begin{figure}[!hbt]
\begin{centering}
\includegraphics[scale=0.8]{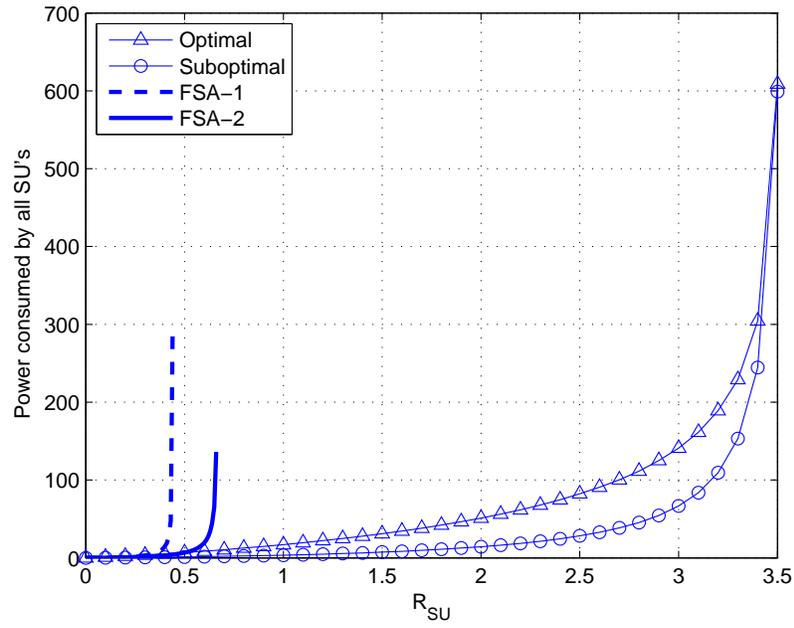}
\vspace{-1em}
 \caption{Average power consumption by all SU's versus $R_{SU}$ at total transmit SNR of 30dB.} \label{fig:2}
\end{centering}
\vspace{-0.1em}
\end{figure}

\begin{figure}[!htb]
\begin{centering}
\includegraphics[scale=0.8]{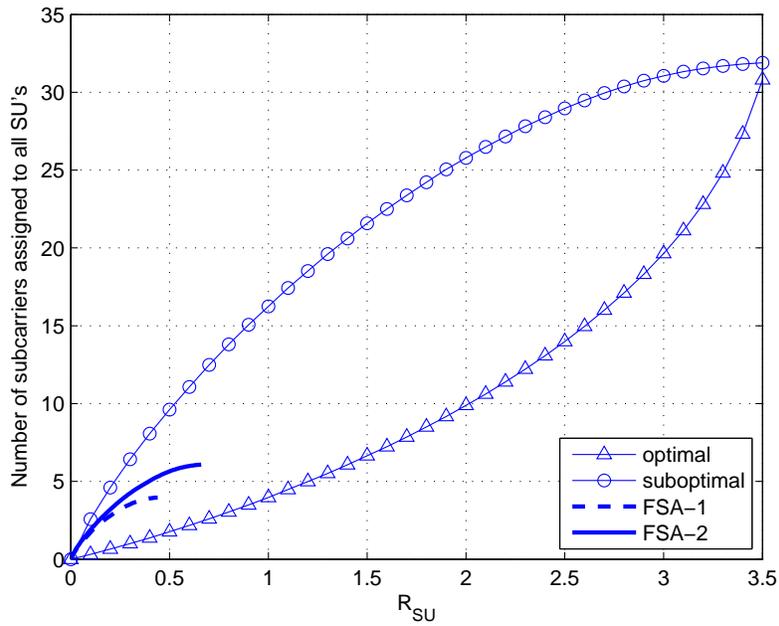}
\vspace{-1em}
 \caption{Average number of subcarriers assigned to all SU's versus $R_{SU}$ at total transmit SNR of 30dB.} \label{fig:3}
\end{centering}
\vspace{-0.1em}
\end{figure}

Fig.~\ref{fig:2} and Fig.~\ref{fig:3} show, respectively, the
average power consumption and the average number of subcarriers
assigned to all SU's with respect to different $R_{SU}$ when the
total average transmit power is fixed as $30$ dB.
From Fig.~\ref{fig:2}, we notice that the optimal scheme spends more
power on SU's than the suboptimal one.
It is observed from Fig.~\ref{fig:3} that the number of occupied
subcarriers by SU's increases with the growing of $R_{SU}$ and
reaches $32$ at the feasible point of $R_{SU}=3.5$ nat/OFDM symbol
for both the optimal and suboptimal schemes. Note that $32$ is an
expected number because when $R_{SU}$ is very close to the feasible
point, adaptive schemes tend to assign as many subcarriers as
possible to SU's and on average, all SU's can occupy half of total
subcarriers at most. Additionally, the optimal scheme assigns less
subcarriers to SU's than the suboptimal one. For FSA-1 and FSA-2,
the number of occupied subcarriers by SU's is also increasing with
the feasible $R_{SU}$ and is smaller than the number of pre-assigned
subcarriers to SU's, i.e. 8 and 12, respectively for FSA-1 and
FSA-2. This indicates that fixed subcarrier assignments waste
subcarriers compared with adaptive ones. Note that the number of
subcarriers assigned to all NU's for the optimal and suboptimal
schemes can be straightforwardly obtained by subtracting the numbers
shown in Fig.~\ref{fig:3} from the total number of subcarriers,
$64$.

\begin{figure}[htbp]
\begin{centering}
\includegraphics[scale=0.8]{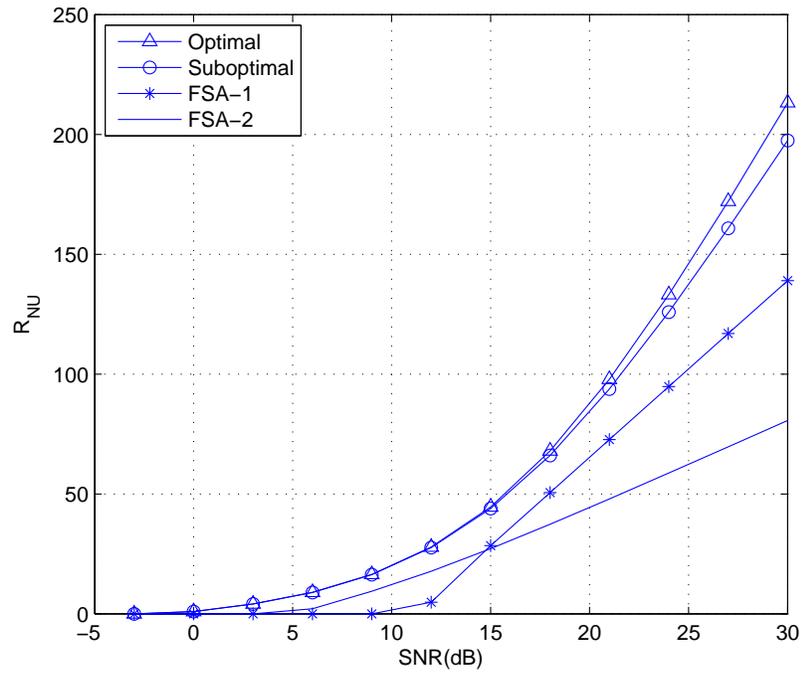}
\vspace{-1em}
 \caption{$R_{NU}$ versus total transmit SNR at $R_{SU}=0.4$nats/OFDM symbol.} \label{fig:4}
\end{centering}
\vspace{-0.1em}
\end{figure}

We next demonstrate the relation between $R_{NU}$ and total transmit SNR
for a given $R_{SU}=$0.4 nat/OFDM symbol in Fig.~\ref{fig:4}. Note
that the constraint $R_{SU} =$ 0.4 is feasible when SNR$\geq$ -2dB for the
optimal and suboptimal schemes and when SNR$\geq$ 3 or 9 dB,
respectively for FSA-1 and FSA-2. From Fig.~\ref{fig:4}, we first observe
that at low SNR region (SNR$\leq$18dB), the optimal and suboptimal schemes
perform close to each other. As SNR becomes larger, the suboptimal scheme
only incurs a marginal performance loss.
It is also seen that the function curves of FSA-1 and FSA-2 intersect
at about 14dB. When the transmit SNR is lower than 14dB, FSA-2 is superior
to FSA-1 in terms of $R_{NU}$. This is because when power is limited, FSA-2
assigns SU's more subcarriers and thus save more power for NU's. If transmit
SNR is higher than 14dB, FSA-2 becomes inferior to FSA-1. The reason lies
in that when power is sufficient to meet secrecy rate requirements, FSA-2
wastes subcarriers on SU's and leaves NU's less to promote their information rates.

\begin{figure}[!htb]
\begin{centering}
\includegraphics[scale=0.8]{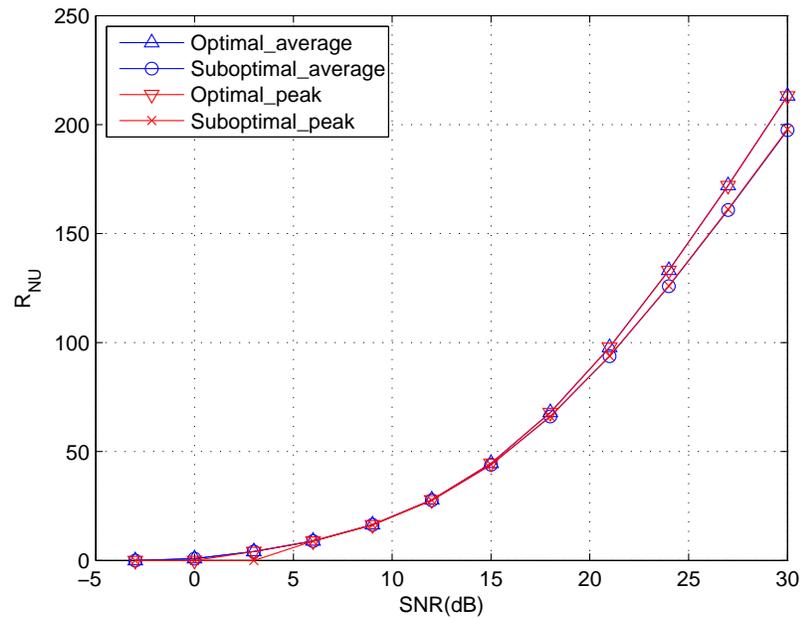}
\vspace{-1em}
 \caption{$R_{NU}$ versus total transmit SNR at $R_{SU}=$ 0.4nat/OFDM symbol under both average and peak power constraints.} \label{fig:5}
\end{centering}
\vspace{-0.1em}
\end{figure}

Finally, in Fig.~\ref{fig:5}, we compare the performance under average and
peak power constraints and show the achieved aggregate rate of NU's at different
SNR. It is observed that under the two power
constraints, the two curves differ slightly in low SNR region and almost
coincide at high SNR region for both optimal and suboptimal schemes.

To conclude the above results, the proposed optimal and suboptimal resource
allocation schemes significantly outperform those with fixed subcarrier
assignment. This observation indicates the great importance of carefully
coordinating the subcarrier allocation with adaptation to the system channel
conditions. Moreover, the suboptimal scheme provides a good tradeoff between
performance and complexity.

\section{Conclusions and Discussions}
This work investigate the power and subcarrier allocation policy
for OFDMA broadband networks where both SU's and NU's coexist. We
formulate the problem as maximizing the average aggregate
information rate of NU's while satisfying the basic average secrecy
rate requirements of SU's under either an average or a peak transmit
power constraint. We solved the problem asymptotically in dual
domain by using decomposition method. Results show that the optimal
power allocation for an SU depends on both its channel gain and the
largest channel gain among others. We also observe that an SU
becomes a valid candidate competing for a subcarrier only if its CNR
on this subcarrier is the largest among all and larger enough than
the second largest CNR. To reduce the computational cost, a
suboptimal scheme with favorable performance is presented. Numerical
results show that the optimal power and subcarrier allocation
algorithm effectively boosts the average total information rate of
NU's while meeting the basic secrecy rate requirements of SU's. It
is also shown that whether it is peak or average power constraint,
the system performance does not differ much given sufficient number
of subcarriers used.

{\color{blue} Before finishing the paper, we provide some discussions.
Firstly, we assumed throughout this work that the channel state information of each user obtained at the BS is accurate. If a user deliberately lies and reports a lower channel-to-noise ratio on certain subcarriers, it would
get higher chance of eavesdropping SU's private message and therefore cause secrecy
rate loss. However, on the other side, its own average information rate or secrecy rate would also be reduced due to the less assigned radio resources. Hence, we argue that there is no incentive for the users to lie about their channel condition. If, however, the network contains purely
malicious eavesdropper that does not care about its own transmission, the security in the network can be circumvented by sending the BS false channel measurements.
Secondly, if eavesdroppers can collude and exchange outputs to decode the message, the network can be regarded as a single-input multiple-output system, where there is only a single eavesdropper with multiple receive antennas. In this case, our algorithms can still apply except the change that the secrecy rate of an SU depends on the sum of the channel gains of all the eavesdroppers rather than the largest one. 
Furthermore, in the case where a user is equipped with multiple receive antennas, an interesting topic for future investigation is to allow the user to report only one antenna to the BS and use the remaining antennas to eavesdrop. 
Lastly, since the proposed resource allocation algorithm is a centralized
one, full knowledge of CSI is required as commonly assumed in the literature. Taking into account practicality
and system overhead, we will investigate distributed algorithms with
partial or local channel knowledge in our future work.}

\bibliographystyle{IEEEtran}

\bibliography{reference}

%
%
%
%

\end{document}